# Toward Responsible ASR for African American English Speakers: A Scoping Review of Bias and Equity in Speech Technology

Jay L. Cunningham[1], Adinawa Adjagbodjou[3], Jeffrey Basoah[2], Jainaba Jawara[5], Kowe Kadoma[4], Aaleyah Lewis [2]

[1]DePaul University — Chicago, IL USA
[2]University of Washington — Seattle, WA USA
[3] Carnegie Mellon University — Pittsburgh, PA USA
[4] Cornell Tech University — New York City, NY USA
[5] University of Maryland — College Park, MD USA

jcunni37@depaul.edu, aadjagbo@andrew.cmu.edu, jeffkb28@uw.edu, jjawara@umd.edu, kk696@cornell.edu, alewis9@cs.washington.edu

**Abstract**

This scoping literature review examines how fairness, bias, and equity are conceptualized and operationalized in Automatic Speech Recognition (ASR) and adjacent speech and language technologies (SLT) for African American English (AAE) speakers and other linguistically diverse communities. Drawing from 44 peer-reviewed publications across Human-Computer Interaction (HCI), Machine Learning/Natural Language Processing (ML/NLP), and Sociolinguistics, we identify four major areas of inquiry: (1) how researchers understand ASR-related harms; (2) inclusive data practices spanning collection, curation, annotation, and model training; (3) methodological and theoretical approaches to linguistic inclusion; and (4) emerging practices and design recommendations for more equitable systems. While technical fairness interventions are growing, our review highlights a critical gap in governance-centered approaches that foreground community agency, linguistic justice, and participatory accountability. We propose a governance-centered ASR life-cycle as an emergent interdisciplinary framework for responsible ASR development and offer implications for researchers, practitioners, and policymakers seeking to address language marginalization in speech AI systems.

# Introduction

The rapid growth and widespread application of Automatic Speech Recognition (ASR) systems have opened new possibilities across domains such as transcription services, accessibility tools, and virtual assistants, promising greater access to users. However, these systems continue to demonstrate systemic disparities in performance, particularly for speakers of African American English (AAE), whose linguistic patterns are frequently misrecognized, misinterpreted, or erased by ASR systems predominantly trained on standardized American English (SAE) datasets (Koenecke et al. 2020; Mengesha et al. 2021). These failures are not merely technical errors; they reflect and reinforce broader societal inequities around whose voices are recognized, valued, and legitimized by AI systems (Wenzel and Kaufman 2023; Brewer, Harrington, and Heldreth 2023; Hanna et al. 2020; Smith et al. 2020).

These harms also reflect long-standing language ideologies that shape which forms of speech are deemed legitimate in society. Language variation, has long been a focus in sociolinguistics, which has shown how dominant attitudes toward non-standard dialects and vernaculars uphold social hierarchies and racialized perceptions. In particular, sociolinguistic scholarship has investigated how language upholds social inequalities and perpetuates bias specifically for AAE speakers (Labov 1963; Rosa and Flores 2017; Morgan 2002; Green 2002a) despite decades of significant sociolinguistic efforts to discredit discriminatory narratives and work to legitimize it as a systematic, rule-governed variety (Labov 1963; Poplack 1980). These views, rooted in deficit-based ideologies (Rosa and Flores 2017; Rickford 2016), persist in language technologies, where users who do not conform to dominant language practices (Blodgett et al. 2020) often face negative consequences.

For AAE speakers, ASR system breakdowns reproduce historical patterns of exclusion, erasure, and discrimination embedded in broader algorithmic and data-driven harms (Wenzel et al. 2023; Bird 2024). These harms manifest as usability failures, trust erosion, and miscommunication, but also extend into moderation, surveillance, and policing contexts, where mis-recognition can have direct material and safety consequences (Mengesha et al. 2021; Cunningham et al. 2024). These systemic harms highlight the urgent need to re-evaluate the ASR development pipeline itself, particularly the universal design assumptions that erase linguistic and cultural specificity.

In this scoping review (Tricco et al. 2018), we focus explicitly on ASR as a speech AI technology, recognizing that while our findings also inform broader Speech and Language Technologies (SLTs) and Natural Language Processing (NLP) systems, the primary site of harm, exclusion, and intervention discussed herein remains ASR systems deployed in U.S. English-language contexts serving Black AAE speakers. Addressing these disparities requires an interdisciplinary synthesis of research practices, data methods, and governance frameworks across HCI, ML/NLP, and so-ciolinguistics, as efforts in these domains have often been fragmented and siloed. While AAE is the focal case for this review, given its sociolinguistic complexity, historical marginalization, and documented ASR disparities, the findings and governance-centered framework are intended to serve as a transferable model for advancing inclusive ASR practices across other underrepresented dialects and languages. This framing ensures the work contributes both to the specific remediation of harms in AAE contexts and to the broader theoretical and practical discourse on linguistic equity in speech AI.

While prior literature reviews, such as Ngueajio and Washington (2022) survey on ASR system biases and mitigation techniques, have synthesized technical approaches to bias identification and correction, these works primarily center model-level interventions and error mitigation strategies. In contrast, this review adopts a broader socio-technical and governance-centered lens, treating ASR fairness not solely as a matter of model optimization but as a systemic issue rooted in data practices, linguistic marginalization, and exclusionary design norms. Our work extends beyond system performance to interrogate whose linguistic practices are privileged, how epistemic authority is distributed in ASR development, and what participatory and community-led models of design and evaluation are emerging or still lacking. By foregrounding linguistic justice, cultural legitimacy, and the governance of ASR systems, this review addresses critical gaps in existing scholarship and contributes to ongoing conversations on decolonial computing and algorithmic harm reduction.

This review seeks to bridge these gaps by synthesizing existing interdisciplinary scholarship, highlighting tensions and synergies across fields, and offering pathways toward more accountable, participatory, and community-centered ASR development.

**Research Questions** This study is guided by the following research questions (RQs):

- RQ1: Across Human-Computer Interaction (HCI), ML/NLP, and Linguistics, how are researchers discussing and addressing fairness, bias, and equity in ASR for Black speakers of AAE?
- RQ2: What are the applied data practices (curation, collection, validation, annotation, deployment) being used or proposed in ASR systems that aim to be inclusive of AAE speakers and Black users?
- RQ3: What broader conceptual design approaches, methods, and theories have been employed to advance inclusive ASR systems for Black AAE speakers?
- RQ4: What are practical recommendations for designing, deploying, or governing more AAE-inclusive ASR systems?

These research questions seek to document the various intervention points where researchers, practitioners, and other stakeholders engage with ASR development, and to assess the tangible outcomes these actions have yielded. Beyond technical practices alone, we take an intentional approach to documenting discourse, practices, and recommendations, with the aim of capturing the expansive strategies that future researchers can build upon to contribute to more accountable, linguistically-inclusive, and community-centered ASR technologies. We focus on AAE as an illustrative case of equity challenges in speech technology, addressing urgent social justice concerns while generating transferable insights for inclusive ASR design across other minoritized dialects and languages.

With this work, we contribute:

1. An interdisciplinary synthesis of the current responsible ASR landscape, critically analyzing how fairness, bias, and equity are addressed across HCI, ML/NLP, and linguistics, with a specific focus on the lived experiences and speech practices of Black AAE speakers.
2. A mapping of data practices, design approaches, and evaluation methods currently used or proposed to mitigate ASR disparities, highlighting tensions, synergies, and disciplinary blind spots.
3. A governance-centered reflection on the limitations of existing approaches, offering pathways toward more participatory, accountable, and community-centered ASR development that explicitly challenges universal design assumptions and centers linguistic justice.

This work advances the AIES agenda by foregrounding the systemic and socio-technical harms embedded in ASR systems, emphasizing that addressing these issues requires not only technical interventions but also structural changes in who governs, designs, and evaluates ASR technologies, and how those processes can be re-imagined to center the voices and agency of Black AAE speakers.

*\*An extended version of this paper, which includes supplementary materials, technical appendices, and expanded methodological details, is available on arXiv[1] under a Creative Commons Attribution (CC BY) license.*

## Background

To situate our scoping review, this section examines how language variation, particularly AAE, intersects with technical infrastructures and social power in ASR. We draw from sociolinguistics, HCI, ML/NLP, and speech engineering to outline the historical and contemporary barriers to inclusive ASR development and position the need for governance-centered intervention.

### Language Variation and Socio-technical Hierarchies

Language is not merely a technical input but a socially encoded practice shaped by ideology, identity, and inequality. Sociolinguistic studies have long demonstrated that linguistic variation across dialects, regions, and communities is both systematic and meaningful (Labov 1963; Green 2002a). Yet, dominant ideologies frame non-standard varieties, especially AAE, as deficient, contributing to racialized hierarchies in education, employment, and increasingly,

---

[1] navigate to https://jaylcunningham.com/work for the arXiv web-link.

AI systems (Rosa and Flores 2017; Rickford 2016). These views persist despite decades of research affirming AAE's grammatical structure and internal consistency (Labov 1972; Poplack 1980; Lanehart 2015).

This foundational work is critical for contextualizing how bias in ASR systems is not an isolated technical artifact, but the continuation of social and linguistic exclusion via algorithmic infrastructures.

## African American English and Inclusive ASR

African American English (AAE), commonly referred to as African American Vernacular English (AAVE) or African American Language (AAL), is a dynamic, rule-governed language variety with deep cultural and historical significance. While early sociolinguistic studies focused narrowly on the speech of urban, working-class Black men, contemporary research demonstrates rich intra-group variation across gender, geography, and age (King 2020; Farrington 2019; Forrest and Wolfram 2019). The fluidity of AAE, shaped by regional, generational, and slang variation, poses challenges for ASR, as standardized training data often miss its shifting phonological, lexical, and stylistic patterns, reducing accuracy and risking intra-community inequities without adaptive, continually updated models (Koenecke et al. 2020; Blodgett, Green, and O'Connor 2016).

Despite this linguistic legitimacy, AAE continues to be stigmatized in language technologies. Studies across HCI, NLP, and ASR consistently show that AAE speakers face higher word error rates, reduced system reliability, and mis-recognition of culturally specific language (Koenecke et al. 2020; Mengesha et al. 2021; Harrington et al. 2022). These failures are not neutral but reflect socio-technical design decisions that privilege SAE as the norm.

Attempts to improve system performance for AAE have included dialect-specific datasets (Blodgett, Green, and O'Connor 2016), syntactic disambiguation tools (Santiago et al. 2022), race-primed annotation protocols (Sap et al. 2019), and participatory design methods (Kim et al. 2022). While these interventions are valuable, they often emerge within siloed disciplines, limiting the field's ability to enact structural change.

## Bias and Disciplinary Gaps in ASR Research

Technical disciplines such as ML and speech engineering have made progress on bias mitigation, data curation, and fairness audits (Hutiri and Ding 2022; Xu et al. 2021), but often lack engagement with the sociocultural dynamics that shape language marginalization. Conversely, HCI and sociolinguistic studies center user perspectives, cultural context, and community agency, but are underrepresented in mainstream ASR pipelines.

Recent studies have surfaced racial bias and underrepresentation in ASR systems (Ngueajio and Washington 2022; Blodgett et al. 2020), yet few bridge insights across disciplines. This fragmentation obscures shared goals and undermines opportunities for holistic, equity-centered innovation. Our review addresses this gap by synthesizing interdisciplinary approaches to bias, equity, and fairness in ASR systems for AAE speakers.

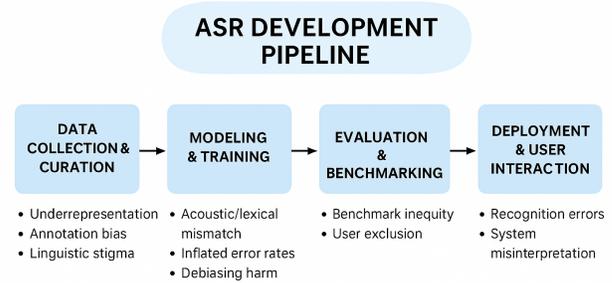

Figure 1: ASR Modeling Development Pipeline and Sites of Harm

## ASR Development Pipeline and Sites of Harm

ASR systems function through a multi-step pipeline, data collection, annotation, modeling, evaluation, and deployment, each of which encodes socio-technical values (see Figure 1). At every stage, AAE speakers are vulnerable to exclusion, mis-recognition, or epistemic erasure.

- **Data Collection & Curation**: Corpora often underrepresent AAE or homogenize its variation, leading to under-performing models and biased assumptions (Dorn 2019; Markl and McNulty 2022).

- **Annotation**: Annotators unfamiliar with AAE may mislabel nonstandard forms or reinforce toxicity stereotypes (Sap et al. 2019; Davidson, Bhattacharya, and Weber 2019).

- **Model Training**: Standard acoustic and language models penalize features common in AAE, resulting in elevated error rates and interpretive failures (Koenecke et al. 2020; Okpala et al. 2022).

- **Evaluation**: Benchmarks calibrated to SAE ignore dialectal differences, erasing AAE-specific challenges from performance metrics (Wenzel et al. 2023).

- **Deployment**: Failures in high-stakes domains like healthcare or employment magnify harm and diminish user trust (Crawford 2021).

By mapping these harms across the ASR life-cycle, our study reframes ASR not just as a technical pipeline, but as a governance challenge, demanding intervention at multiple sites of design and accountability. ASR's failures for AAE speakers illustrate a broader crisis in language AI: the reproduction of racial and linguistic hierarchies through automation. By articulating the socio-technical dimensions of harm and highlighting underexplored disciplinary gaps, this review invites a more ambitious and justice-centered vision for speech technologies, one that moves beyond representational diversity toward equity-centered governance and design.

## Scoping Review Methodology

This review adopts the PRISMA Scoping Review (PRISMA-ScR) methodology as outlined by (Tricco et al. 2018), which is specifically suited for comprehensively mapping emerging and heterogeneous research fields such as ASR. This review is based solely on secondary analysis of published literature and does not include new empirical testing of ASR systems. Given the interdisciplinary nature of ASR scholarship, this review employs an inductive scoping approach to capture the diversity of epistemological positions, data practices, and design approaches present across disciplines, including HCI, ML/NLP, and sociolinguistics, while remaining open to underrepresented perspectives, such as those centering Black AAE speakers. Scoping reviews are particularly appropriate for fields where the research landscape is complex, fragmented, and under-reviewed, as they provide a structured yet exploratory assessment of the extent, range, and nature of research activity. By applying this method, we aim to map and synthesize the interdisciplinary discourse on ASR fairness and equity, identify knowledge gaps, and inform future research and governance directions.

## Inclusion and Exclusion Criteria

This inclusion and exclusion criteria framework guide our literature review process.

**Inclusion Criteria** Studies are included if they meet any of the following conditions: 1) Focus specifically on AAE or similar American English linguistic variations, 2) Have a strong thematic focus on bias identification, inclusive ASR/NLP design methods, and practical recommendations for inclusive design, aimed at improving the fairness and inclusion of ASR technologies.

**Exclusion Criteria** Studies are excluded if they: 1) Do not focus on AAE or similar ethnic minority linguistic variations of American English, 2) Are primarily focused on other languages or general ASR technologies without a clear emphasis on inclusivity, equity, or responsible design practices.

**Information Sources and Search Strategy** To guide our interdisciplinary literature search, we first outlined six thematic clusters reflecting key conceptual areas of fairness and equity in ASR systems: (1) Responsible Methods & Design Approaches, (2) Social & Algorithmic Justice Challenges, (3) Speech & Language Technologies (ASR Focus), (4) Language Varieties & Dialects (U.S. Context), (5) Racial & Ethnic Communities, and (6) Language Group Associations. These clusters were developed to ensure conceptual breadth and to capture both technical and sociolinguistic framings relevant to ASR fairness.

Then, we developed an initial seed set of keywords informed by prior work in AI fairness, sociolinguistics, and speech technology (e.g., "algorithmic bias," "dialect recognition," "African American English," "ASR fairness," "voice equity," "linguistic inclusion"). We then broadened our search terminology using ResearchRabbit's citation network mapping to broaden the conceptual landscape and avoid disciplinary blind spots. This tool generated a network of 1,872 semantically adjacent and related terms, expanding coverage beyond initial expert-defined keywords and surfacing domain-specific variants (e.g., "voice bias mitigation," "vernacular speech tech," "speech dialect equity"). This expansion process ensured coverage across ML/NLP, HCI, and sociolinguistics while capturing emergent terminology [2].

## Data Charting Process

**Extraction of Evidence** Initially, 72 articles were identified and cataloged using Airtable. After applying the inclusion and exclusion criteria, 44 papers were selected for further analysis. These papers specifically focus on inclusive design principles within ASR technologies. The final sample of 44 papers reflects the narrow scope of research explicitly engaging with fairness, bias, or equity in ASR for African American English or comparable non-standard American English varieties. While the number may appear small, it underscores the limited scholarly attention to ASR equity for marginalized speech communities and the need for targeted research in this area.

**Charting of Evidence**[3] The selected 44 papers were processed using Atlas.ti, a qualitative data analysis software, to facilitate coding. A preliminary codebook was created based on the research questions, employing open coding for emergent themes related to bias detection, inclusive design strategies, and diverse linguistic representation (see Figure 2). This systematic approach allowed for a structured analysis of the collected data, ensuring that all relevant themes were thoroughly explored and documented.

**Disciplinary Mapping** To assess the interdisciplinary landscape of fairness and equity in ASR systems, we classified the 44 reviewed papers by their primary disciplinary orientation based on publication venue, author expertise, and topical focus. The majority of papers (n=31) were situated within Machine Learning and Natural Language Processing (ML/NLP), reflecting the technical emphasis on bias mitigation, data practices, and model evaluation in speech and language technologies. Human-Computer Interaction (HCI) contributed 11 papers, centering on user experience, participatory design, and socio-technical impacts of ASR systems, particularly for marginalized communities. Linguistics was represented in 2 papers, focusing on the sociolinguistic characteristics of AAE and their implications for ASR performance. This distribution underscores the predominance of technical approaches in the current literature, while highlighting emerging intersections with HCI and linguistics that offer critical perspectives on ASR equity.

**Data Items** The following details were charted for each article where applicable: Author(s); Year of publication and venue (e.g. CHI, 2023); Paper Title; Origin/country of origin (where the study was published or conducted); Aims/purpose and research questions; Study population or applica-

---

[2] (See the Appendices (B) in extended paper version for expanded details of representative terms for each cluster.)

[3] (See the Appendices (B) in extended paper version for the full table of Reviewed Literature organized by Citation, Discipline, and Information Source.)

bility to ethnic and linguistic minority American English speaking groups; The type of study (conceptual or empirical); Methodology

**Data Analysis Team** The data was coded by a team of five reviewers, all of whom are experienced researchers with diverse backgrounds in HCI, Computer Science (AI/ML), NLP, and Linguistics (Computational and Social). This diversity in expertise enhances the reliability and breadth of the analysis.

We employed an inductive coding approach using Atlas.ti, guided but not constrained by the research questions. A collaboratively developed codebook emerged through iterative cycles of reading and open coding across the full texts of included papers. Two authors independently coded each paper in each round of reviews, and inter-coder agreement was calculated using *pairwise* Cohen's kappa (Cohen 1960), averaged across all coder pairs, at each iteration. With six coders annotating the full set ($N = 44$ items), this involved computing $\kappa$ for each of the $\binom{6}{2} = 15$ coder pairs and taking the mean. Inter-coder reliability for the initial round of full-text coding was $\bar{\kappa} = 0.82$, indicating strong agreement. Disagreements were resolved through group discussion, and the final codebook was applied with 100% consensus across all included papers. The resulting thematic structure (see Section reflects both the empirical diversity of reviewed works and the normative framing of responsible ASR design embedded in our Research Questions.

**Synthesizing of Results** The synthesis process aggregated the findings from these studies, identifying common patterns and themes as well as gaps in the research. This analysis helped in understanding how responsible design practices are conceptualized and implemented in ASR systems for speakers of AAE, identifying key areas for future research.

**Summary of Data** The table below shows the number of documents retrieved from various information sources (see Table 1)

Table 1: Summary Search Protocol

| Elements | Details (count) |
| --- | --- |
| Search Keywords | "Algorithmic Fairness" |
| Limit To | Scholarly articles (peer-reviewed or otherwise) |
| ACLWeb | 18 |
| ACM Digital Library | 19 |
| IEEE Xplore | 4 |
| arXiv | 1 |
| PNAS (National Academies) | 1 |
| Frontiers | 1 |
| ProQuest | 1 |

## Scoping Review Summary and Findings

Following PRISMA-ScR (Tricco et al. 2018) for the reporting of scoping review results, we first present the process of selecting sources of evidence. We then describe the characteristics of these sources of evidence in terms of publication venue and year. And finally, we present the synthesis of the results answering each research question.

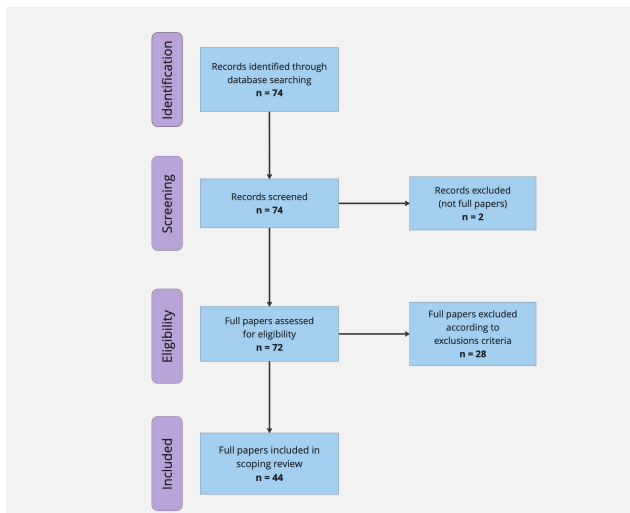

Figure 2: Scoping Review Flow Diagram following the PRISMA method for scoping reviews.

## Framing Fairness and Harm in ASR: Disciplinary Perspectives from HCI, ML/NLP, and Linguistics

This section addresses Research Question 1. We organize findings by disciplinary emphasis, highlighting how each field frames fairness, bias, and equity, and synthesizing their distinct contributions and blind spots.

**ML/NLP: Quantifying Disparities and Model Fairness** In ML/NLP, researchers primarily approach ASR fairness through performance evaluation and bias mitigation metrics. Koenecke et al. (2020) document racial disparities in ASR word error rates using commercial systems, revealing that systems perform disproportionately poorly for Black speakers. Sarı, Hasegawa-Johnson, and Yoo (2021) propose counterfactual fairness interventions in ASR models, but do not explicitly account for dialectal variance in AAE. Xu et al. (2021) and Ziems et al. (2022) demonstrate that dialect disparity extends beyond ASR into language understanding tasks, showing how minority dialects are marginalized in pretraining and classification phases. Manzini et al. (2019) and Davidson, Bhattacharya, and Weber (2019) uncover bias in hate speech datasets and embeddings, often embedding stereotypes that affect downstream fairness.

Other works explore fairness in speech data augmentation or annotation. Johnson et al. (2022) propose LPC-based augmentation to support underrepresented children's dialects, while Okpala et al. (2022) use adversarial training to debias hate speech detection. However, despite advances in model-based solutions, few works in this domain consider participatory definitions of fairness or sociolinguistic legitimacy of dialectal inputs. Fairness is often operationalized via demographic parity or performance parity, without scrutiny of epistemic assumptions or linguistic ideologies embedded in models.

**HCI: Trust, Experience, and Interaction Harms** HCI studies frame ASR fairness through trust, user experience, and emotional impact. Wenzel et al. (2023) and Wenzel and Kaufman (2023) find that ASR , mis-recognition can function as a microaggression, especially across race-aligned interactions. Harrington et al. (2022) examine older Black adults' health information-seeking behaviors, finding that ASR failures prompt code-switching, self-silencing, and distrust. These findings emphasize the affective burden and social exclusion that arise from ASR unreliability.

Accessible interface work such as Lister et al. (2020) and Kim et al. (2022) proposes inclusive design strategies, but does not often evaluate performance differentials along race or dialect lines. Cunningham (2023) calls for community-collaborative approaches to ASR development with AAE speakers to address these lived harms. However, most HCI research does not quantify disparities or directly influence model-level design decisions, indicating a methodological disconnect between user experience research and ML fairness interventions.

**Linguistics and Sociolinguistics: Structural Marginalization and Language Ideology** Linguists and sociolinguists emphasize how ASR systems encode structural linguistic inequities. Blodgett, Green, and O'Connor (2016) and Rickford (2016) argue that AAE is systematically devalued in both NLP and speech datasets. Bird (2024) critiques the colonial legacies of linguistic extraction, warning that even fairness-oriented inclusion can reinforce linguistic imperialism. Bourdieu's (1991) theories of linguistic capital are used by Weston (2021) and Rosa and Flores (2017) to show how ASR systems reinforce dominant language ideologies.

Recent work by Markl and McNulty (2022) and Markl (2022) highlights how ASR practitioners act as language managers, often enforcing standard language norms while overlooking dialect-specific error patterns. Wassink, Gansen, and Bartholomew (2022) examine acoustic alignment failures with ethnicized dialects in ASR, revealing mismatches in phonetic modeling. Santiago et al. (2022) propose improved modeling of morpho-syntactic features in AAE such as habitual "be," arguing for dialect-sensitive annotation and modeling.

**Cross-Cutting Synthesis: Toward Intersectional and Participatory Fairness** Across disciplines, researchers converge on the finding that ASR systems disproportionately fail minoritized speakers, particularly those using AAE, through high error rates, coerced linguistic assimilation, and erasure of dialectal legitimacy. Yet, few studies bridge technical, interaction, and ideological dimensions. ML/NLP largely quantifies disparities; HCI explores experience and harm; linguistics interrogates ideological assumptions. Interdisciplinary synthesis remains limited.

Notably, participatory approaches are largely absent. While some HCI and sociolinguistic studies advocate for community-centered design (Cunningham 2023; Harrington et al. 2022), no papers in ML/NLP implement participatory fairness definitions or community-led audits. This gap suggests a critical opportunity for future research: integrating intersectional, participatory, and governance-aware practices across the full ASR development life-cycle.

## Data Practices for Inclusive ASR Design

This section addresses Research Question Two. We analyze reviewed studies through four major stages of the data life-cycle: collection, curation, annotation, and model training.

**Data Collection and Community Participation** Inclusive data collection practices for AAE vary across domains, from scraping social media corpora to crowdsourcing speech. In NLP, researchers have collected text from Twitter and other user-generated platforms to compile AAE datasets (Groenwold et al. 2020; Blodgett, Green, and O'Connor 2016; Jørgensen, Hovy, and Søgaard 2015). These datasets often provide linguistic diversity for fine-tuning models but risk entrenching biases if demographic provenance is unclear.

Speech-based collection efforts involve more participatory methods. Markl and McNulty (2022) and Nkemelu et al. (2023) describe community-driven and paid-volunteer approaches to low-resource speech data collection. These methods not only increase linguistic representation but foster community agency in shaping how data is interpreted, especially for hate speech and sociocultural content. However, concerns remain around consent, data commodification, and ownership in crowdsourced platforms.

**Curation and Dataset Integration** Responsible data curation practices appear across HCI and NLP. Researchers curate corpora that document AAE variation, such as the Corpus of Regional African American Language (CORAAL) (Thomason 2021) and IViE (Grabe, Nolan, and Farrar 1998), which have been used in downstream ASR systems (Farrington 2019). These curated datasets feature transparent regional tagging and transcription practices, facilitating dialect-aware modeling.

Some works combine heterogeneous datasets to augment underrepresented forms of AAE. For example, Santiago et al. (2022) and Johnson et al. (2022) integrate multiple sources for robust modeling. Others apply pre-processing methods like punctuation stripping (Dorn 2019) or more complex adversarial auto-encoding for debiasing (Sarı, Hasegawa-Johnson, and Yoo 2021). Manzini et al. (2019) propose a post-curation debiasing step that removes analogical bias from word embeddings before training. These methods reflect growing interest in integrating fairness into data preparation, though sociolinguistic validation is often limited.

**Annotation and Fairness Sensitivity** Annotation is a key site where bias and subjectivity enter ASR pipelines. Bender and Friedman (2018) and Poletto et al. (2021) argue for documenting annotator demographics and training to mitigate interpretive bias. Several studies report using AAE speakers as annotators to assess or validate interpretations of AAE texts and speech. Groenwold et al. (2020) and Sap et al. (2019) show that race-primed annotators are less likely to label AAE tweets as offensive, suggesting that cultural and racial alignment improves interpretive fairness.

Participatory annotation remains rare but promising. Community-informed schema or iterative consensus models could support more epistemically just annotation practices, especially when documenting sociolinguistic nuance or culturally grounded intent.

**Training and Dialect-Specific Modeling** Inclusive ASR development also requires training practices attentive to dialectal features. Okpala et al. (2022) retrain BERT on over one million AAE tweets to produce AAEBERT, a hate speech detection model sensitive to AAE syntax and usage. Similarly, Santiago et al. (2022) propose a framework to disambiguate habitual vs. non-habitual uses of "be" in AAE using rule-based filters, while Ziems et al. (2022) construct orthographic variants through morphosyntactic transformation.

Such techniques show promise for recognizing linguistic specificity, but raise questions about over-generalization and ethical representation. Models trained with dialectal precision may still misinterpret intent or meaning without contextual grounding. Few of the reviewed studies integrate human feedback loops or community validation post-training.

Overall, across the data life-cycle, efforts to include AAE speakers in ASR development remain fragmented. While representational improvements are emerging in collection and modeling, participatory governance, community consent, and epistemic alignment remain underdeveloped. This suggests a need for socio-technically integrated data practices which embed linguistic, cultural, and ethical considerations throughout the pipeline.

## Approaches to Studying Inclusive ASR and NLP

This section addresses Research Question Three. We categorize the reviewed literature into three overarching methodological paradigms: quantitative, qualitative, and theoretical.

**Quantitative Approaches: Measuring Disparity and Model Intervention** Quantitative approaches dominate the reviewed literature, often aiming to measure or mitigate disparities through statistical or computational means. Core methods include word error rate (WER) analysis, comparative classifier evaluation, sentiment scoring, and debiasing.

Several studies quantify model-level harms: Koenecke et al. (2020), Dorn (2019), and Johnson et al. (2022) use WER to expose transcription disparities between AAE and SAE, while Markl and McNulty (2022) employs regression analysis to model algorithmic bias in British English ASR. Wenzel et al. (2023) and Davidson, Bhattacharya, and Weber (2019) use statistical analysis to assess psychological effects and classification bias. Groenwold et al. (2020) conducts BLEU/ROUGE comparisons and sentiment analysis on GPT-2 outputs for AAE and SAE.

Debiasing and detoxification methods offer technical mitigation pathways. Xu et al. (2021) explores four strategies to guide generative models away from toxic outputs, while Manzini et al. (2019) proposes a multiclass embedding debiasing algorithm. Framework-guided quantitative work includes Joshi et al. (2020), who builds a language resource disparity taxonomy, and Hutiri and Ding (2022), who proposes a bias quantification framework for speaker verification challenges.

While these approaches offer reproducible measures of disparity, they often overlook community-defined fairness goals and fail to interrogate how inclusion is conceptually defined.

**Qualitative and Participatory Approaches: Centering Lived Experience and socio-technical Context** Qualitative studies investigate inclusive ASR through user-centered and community-engaged lenses. Common methods include interviews (Kim et al. 2022), surveys and diary studies (Mengesha et al. 2021; Wenzel et al. 2023), and observation (Iacobelli and Cassell 2007). These approaches capture affective, psychological, and sociocultural responses to ASR failures, particularly among Black and AAE-speaking users.

Participatory design methods, spanning co-design (Kim et al. 2022), community-based design (Racadio, Rose, and Kolko 2014), and cooperative inquiry (Björgvinsson, Ehn, and Hillgren 2010), emerge across several HCI-oriented studies. Ziems et al. (2022) and Nkemelu et al. (2023) apply expert-informed validation protocols and context-sensitive annotation, respectively, to guide benchmark or model development. Others use participatory annotation strategies to better preserve dialectal meaning or assess offensive content interpretation (Groenwold et al. 2020; Sap et al. 2019; Jørgensen, Hovy, and Søgaard 2015).

These methods not only enhance design inclusion but function as socio-technical governance mechanisms, redefining who participates in NLP knowledge production and how authority is distributed. Yet, most remain confined to design phases and are rarely integrated into upstream model development or downstream evaluation.

**Theoretical Frameworks: Critical Lenses on Power, Language, and Identity** A smaller subset of studies ground their analysis in theory-driven inquiry. Markl and McNulty (2022) applies intersectionality to trace compounding harms across racial, linguistic, and geographic dimensions. Sutton et al. (2019) introduce sociophonetic-inspired design strategies that highlight context-awareness and voice individualization. Bird (2024) adopts a decolonial approach to contest extractive language technologies and advocate for Indigenous sovereignty over linguistic data.

Frameworks such as microaggressions and harm-based equity (Wenzel et al. 2023), and Critical Race Theory (Crenshaw 1991), guide ethical critique and propose reparative justice principles. These works surface epistemic injustice in NLP and call for more reflexive and politicized design practices. However, such frameworks are rarely operationalized into data, modeling, or evaluation standards.

In sum, while diverse methodological traditions are represented in the literature, most approaches remain siloed. Quantitative studies dominate, but often neglect cultural nuance or participatory principles. Qualitative and theoretical work illuminate structural harms and lived experience but lack integration into dominant AI pipelines. Future research must advance mixed-method and theory-informed approaches that foreground the agency of marginalized speakers in defining and achieving fairness.

## Practical Recommendations for Inclusive ASR/NLP Design

To explore Research Question Four, this section synthesizes practical strategies for ASR development.

**Community Engagement and Problem Framing** Designing inclusive systems begins with centering communities in defining problems, goals, and design values. Research in HCI and sociolinguistics underscores the need to co-develop technology with the linguistic communities it is meant to serve (Kim et al. 2022; Cunningham 2023; Bird 2024). These studies engage AAE speakers through participatory design and interviews, surfacing culturally relevant use cases, frustrations with existing systems, and desired features. Scholars advocate reframing speech tech development as civic infrastructure (Markl and McNulty 2022), where public input, trust, and ethical alignment guide innovation from the outset (Mengesha et al. 2021; Brewer, Harrington, and Heldreth 2023).

**Ethical and Inclusive Data Collection** Inclusive ASR/NLP development depends on intentional, consent-based data collection that reflects community norms and linguistic realities. Researchers emphasize collecting speech data from diverse AAE speakers using both social media corpora (Groenwold et al. 2020; Blodgett, Green, and O'Connor 2016) and community-centered efforts like Project Elevating Black Voices, a partnership where data sovereignty remains with historically Black institutions (HBCUs) (Heldreth 2023). In low-resource contexts, community-led data practices, such as those by Māori speakers or hate speech annotators in Sub-Saharan Africa, provide models for ethical stewardship (Nkemelu et al. 2023; Brown et al. 2024).

**Responsible Data Curation and Annotation** Annotation must be approached as a socio-technical process shaped by language ideologies and annotator positionality (Bender and Friedman 2018; Papakyriakopoulos and Xiang 2023). AAE-inclusive projects often recruit code-switching annotators or apply dialect/race priming to mitigate labeling bias (Sap et al. 2019; Groenwold et al. 2020). Situated annotation practices, matching annotators to speaker demographics, help preserve meaning and reduce misidentification (Papakyriakopoulos and Xiang 2023). These approaches demand transparency around annotator training and a commitment to dialectal accuracy.

**Dataset Design Dynamics and Representation** Inclusive datasets go beyond quantity to account for representational ethics, balance, and diversity (Jo and Gebru 2020). Scholars propose curating datasets by dialect, region, and sociocultural attributes (Mengesha et al. 2021; Markl and McNulty 2022), and balancing linguistic data across sociodemographic strata (e.g., age, gender, socioeconomic status) (Wassink, Gansen, and Bartholomew 2022). Assets-based approaches shift the design lens from deficit to strength, affirming the linguistic value of marginalized varieties (Wenzel et al. 2023; Harrington et al. 2022).

**Inclusive Model Training and Evaluation** Model development must accommodate linguistic variation while protecting communities from harm. Techniques such as dialect-specific pretraining (e.g., AAEBERT) (Okpala et al. 2022), rule-based transformations (Ziems et al. 2022), and adversarial filtering (Sarı, Hasegawa-Johnson, and Yoo 2021) offer concrete practices for mitigating dialectal bias. Scholars argue that WER-based metrics must be augmented with qualitative evaluations, including microaggression impacts and user trust (Koenecke et al. 2020; Wenzel et al. 2023). Others propose dynamic evaluation thresholds by speaker group to minimize predictive harms (Hutiri and Ding 2022).

**Deployment, Customization, and Accountability** Deployment strategies should promote user autonomy and system responsiveness. Context-aware and individualized designs allow users to select preferred voices and correct misrecognitions (Sutton et al. 2019; Mengesha et al. 2021). ASR systems can be designed to issue affirmations or apologies following recognition failures, redistributing accountability to the system rather than the user (Wenzel et al. 2023). Researchers propose community-based review panels or impact assessments to ensure ethical use in high-stakes contexts (e.g., health, education, policing) (Blodgett et al. 2022).

**Reflexive and Interdisciplinary Evaluation** Finally, inclusive ASR/NLP development must incorporate mechanisms for continuous reflection and adaptation. This includes positionality statements (Allen 1996), diversity audits, and community accountability protocols (Joshi et al. 2020; Sutton et al. 2019). Sociolinguistics and raciolinguistics provide critical tools for interpreting model behavior and social impact (Rickford 2016; Wenzel and Kaufman 2023). Future work should build interdisciplinary teams and embed equity goals throughout project governance, from funding to peer review.

In sum, the inclusive ASR/NLP life-cycle framework outlined here moves beyond piecemeal fixes to articulate a governance-centered, community-rooted agenda for language technology. These interventions are not just technical, they are socio-technical, epistemic, and political. By reimagining development through these lenses, this review offers scholars and practitioners a scaffold for building systems that affirm, rather than erase, the linguistic richness of marginalized communities.

## Discussion

This scoping review analyzed 44 interdisciplinary publications across ML/NLP, HCI, and sociolinguistics to assess how fairness, bias, and equity are addressed in ASR systems for Black speakers of AAE. Rather than applying a pre-existing analytic lens, we adopted an inductive synthesis approach, allowing critical insights to emerge through comparative literature analysis. One striking insight is the absence of governance-oriented interventions across most of the reviewed literature. While technical fairness auditing and inclusive data practices are present, few studies interrogate how power, accountability, and community agency shape ASR system outcomes. This finding signals a significant gap and opportunity to situate fairness socio-technical commitment to be governed.

## A Framework for Governance-Centered ASR

The findings from our scoping review directly motivate the design of the governance-centered ASR life-cycle. Despite progress in bias detection and data diversification, current fairness interventions in ASR overwhelmingly focus on narrow technical fixes, model tuning, dataset re-balancing, and retrospective evaluations of word error rates (Sap et al. 2019; Koenecke et al. 2020). These interventions are important but insufficient. They often abstract fairness into quantifiable metrics without attending to the underlying political, social, and epistemic power structures that shape how ASR systems are developed, deployed, and governed (Blodgett et al. 2020; Madaio et al. 2020). In response, we propose a Governance-Centered ASR life-cycle as a conceptual and actionable approach of operationalizing responsibility in speech AI. This governance-centered life-cycle is not presented as a separate contribution from the review itself; rather, it distills and integrates the recurring patterns, gaps, and community priorities identified across the surveyed literature into a cohesive framework for future ASR development.

We define a governance-centered ASR life-cycle as a socio-technical framework that embeds community agency, participatory oversight, and institutional accountability across all stages of ASR system development, from problem framing and data collection to evaluation, deployment, and post-deployment auditing. This framework identifies participatory checkpoints across six stages: (1) problem definition, (2) data sourcing, (3) model training, (4) evaluation, (5) deployment, and (6) post-deployment governance. This framework is not merely a sequencing of technical processes but a normative proposal for reorienting ASR development toward inclusive, just, and reflexive practices. Governance here is not simply oversight after deployment, but a continuous set of ethical, political, and participatory commitments embedded across the ASR pipeline [4].

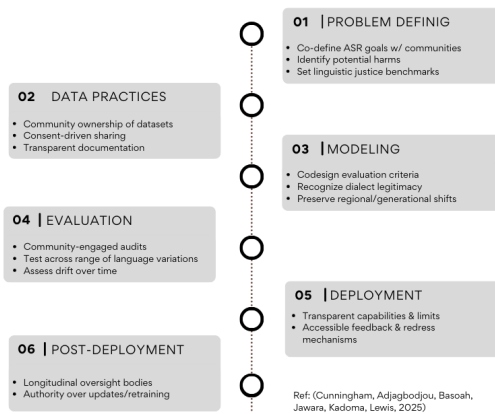

A key design imperative, especially for AAE, is acknowledging its fluidity, regional variation, and rapid incorporation of new lexical and phonological features. ASR systems that fail to account for this dynamism risk obsolescence and cultural erasure, since models trained on static datasets will lag behind evolving usage (Farrington 2019; Green 2002b). A governance-centered approach treats this variability not as a modeling problem to be minimized, but as a feature to be preserved through continuous dataset updates, community-led monitoring, and dialect-sensitive evaluation criteria.

We identify three reasons why governance-centered approaches remain underexplored: (1) interdisciplinary disconnects between ML/NLP, HCI, and sociolinguistics hinder holistic system-level thinking; (2) infrastructural limitations, such as proprietary datasets and black-box model architectures, prevent transparency and shared decision-making; and (3) existing incentive structures in academia and industry often prioritize technical novelty over participatory integrity. These challenges make it difficult to operationalize governance, even when its necessity is acknowledged. We find that few studies operationalize participatory governance at scale. Participatory design and annotation studies remain pilot-scale, often lacking institutional continuity, sustainable funding, or decision-making power shared with communities (Sloane et al. 2020). Where community feedback is gathered, it is too often treated as consultation rather than co-governance. Most industry-led audits and bias evaluations are still conducted internally, without transparency, external validation, or community accountability mechanisms.

The value of a governance-centered life-cycle is substantial. It provides a structured lens through which technologists can anticipate and mitigate downstream harms, assess system legitimacy, and design feedback loops that amplify community voice. It aligns with calls from decolonial computing (Irani et al. 2010; Ali 2016), public interest technology (Eubanks 2018), and AI ethics communities that advocate for accountable socio-technical infrastructures. By embedding participation, consent, and oversight at each stage of the ASR life-cycle, this framework offers a roadmap for equitable and culturally responsive speech AI development.

We argue that a framework for a governance-centered life-cycle requires a structural shift in how ASR systems are built:

- **At the problem definition stage**, governance means defining ASR goals in collaboration with community stakeholders, identifying potential harms, and setting linguistic justice benchmarks before technical work begins.
- **At the data stage**, this means community ownership of speech datasets, consent-driven data sharing, and transparent documentation that centers community goals (Papakyriakopoulos and Xiang 2023; Nkemelu et al. 2023).
- **At the modeling stage**, it entails co-designing evaluation criteria, recognizing dialectal legitimacy, and ensuring linguistic variation, including AAE's regional and generational shifts, is preserved as a design feature rather than normalized away (Ziems et al. 2022; Prinos, Patwari, and Power 2024).
- **At the evaluation stage**, governance includes community-led audits, testing across regional AAE variants, and assessing model performance over time to catch drift in recognition accuracy.

---

[4] (See the Appendices (A) in extended paper version for visual graph of Governance-Centered ASR Life-cycle Framework.)

- **At the deployment stage**, it means establishing transparent communication about system capabilities and limitations, and building accessible user feedback and redress mechanisms.
- **At the post-deployment stage**, governance involves maintaining longitudinal oversight bodies (e.g., community advisory boards, ASR data trusts) with real decision-making authority over updates, retraining, and decommissioning (Blodgett et al. 2022; Ali 2016).

This reframing draws from infrastructure studies (Eubanks 2018), postcolonial computing (Irani et al. 2010; Ali 2016), and recent calls in AI ethics to move from fairness metrics toward democratic accountability (Sloane et al. 2020; Mohamed, Png, and Isaac 2020). It is animated by work showing that community-aligned processes, such as those proposed in Māori ASR initiatives (Brown et al. 2024), Black speech data governance frameworks (Heldreth 2023), and participatory ASR toolkits (Reitmaier et al. 2022, 2023), can produce both culturally competent systems and a deeper trust among impacted communities.

By directly addressing the dynamic, evolving nature of AAE, this approach not only counters the epistemic exploitation and algorithmic erasure identified in critical sociolinguistics (Rosa and Flores 2017), but also ensures that ASR systems remain adaptable to real-world linguistic change. Ultimately, the governance-centered ASR life-cycle advances a justice-oriented vision for speech AI: one where fairness is co-defined, power is redistributed, and community stewardship, not corporate discretion, guides the future of voice-based technologies.

### Cross-Disciplinary Blind Spots

Each disciplinary domain contributes partial perspectives. ML/NLP scholarship prioritizes technical rigor but rarely incorporates community-grounded linguistic expertise (Shah, Schwartz, and Hovy 2020). HCI and sociolinguistic studies foreground user experience and cultural specificity but may lack pathways to influence large-scale system development (Bird 2024; Harrington et al. 2022). The result is a fragmented landscape of interventions. Building a governance-centered life-cycle requires epistemic humility (Hanna et al. 2020) and methodological hybridity that bridges these silos. This echoes broader calls in FAccT and AIES scholarship to transcend disciplinary silos when confronting structural bias (Birhane et al. 2022).

### Participatory Governance as Ethical Imperative

Participatory design and audit mechanisms offer promising pathways for embedding community accountability (Veale and Brass 2018; Cunningham 2023). However, few studies operationalize community co-governance in the evaluation or deployment of ASR systems. Projects like "Elevating Black Voices" (Heldreth 2023) and Māori-led NLP initiatives (Brown et al. 2024) illustrate models of data sovereignty and cultural stewardship that go beyond inclusion toward structural reorientation. Without such models, the risks of epistemic exploitation and linguistic erasure persist (Blodgett et al. 2022). This reorientation invites new research questions. We urge future scholarship to move from inclusive design to inclusive governance, embedding accountability not only in the system, but in the institutions that build and deploy them.

### Study Limitations

In regards to our Methods, we acknowledge that future scoping reviews could benefit from a more systematic documentation of keyword inclusion decisions and filtration pathways, which we identify as a methodological learning from this review process. We also acknowledge that review-writing itself is a form of scholarly gate-keeping (Smith and Dimmick 2010), shaping which works, voices, and epistemologies become canonized in a field where structural inequities already limit whose contributions are visible.

We also recognize the limitations of relying solely on peer-reviewed literature. Our review excludes grey literature, community-authored reports, and archival sources such as Black Web archives. While this decision ensured methodological consistency and comparability across peer-reviewed works, it omits valuable community knowledge and practice-based evidence that may provide further insight into culturally grounded ASR development.

Lastly, we recognize that this review does not include an empirical evaluation of ASR system performance. Future work should pair empirical ASR benchmarking—particularly on African American English and other marginalized varieties—with governance-centered design evaluations to validate and extend the recommendations advanced in this review.

### Conclusion

This scoping review synthesizes how fairness, bias, and equity are addressed in ASR for AAE and presents a framework grounded in linguistic justice, offering transferable lessons for designing ASR systems inclusive of linguistically diverse communities. Using the PRISMA framework, we synthesized 44 interdisciplinary papers across HCI, ML/NLP, and sociolinguistics, identifying four key domains: (1) evolving understandings of ASR harms; (2) inclusive data practices; (3) methodological and theoretical advances; and (4) practical design and governance recommendations.

Our analysis surfaced a persistent gap: while technical interventions abound, few works engage deeply with questions of governance, power, and community agency. In response, we propose a governance-centered ASR life-cycle, a future-oriented framework that re-imagines fairness as a participatory, socio-technical process. We urge future work to advance this direction by co-developing participatory governance models and embedding linguistic justice across the ASR pipeline. Future work should also empirically and collaboratively validate this framework with diverse speech communities by co-developing benchmarks, refining models to community priorities, and evaluating social and technical impacts.

## Ethical Considerations

This review synthesizes interdisciplinary scholarship on ASR, fairness, and linguistic inclusion, with a focus on AAE. We acknowledge that this work is not neutral: it involves interpretive choices that shape which perspectives are elevated in a field where Black voices and linguistic diversity are often marginalized.

Our ethical commitments are rooted in our positionality. (Harding 2004), as we advance care for underrepresented communities, epistemic humility across disciplines, and transparency in how we synthesize and critique scholarship. This work is led by Black researchers whose lived experiences and interdisciplinary expertise in HCI and sociotechnical AI inform a commitment to addressing systemic anti-Blackness and linguistic erasure.

## Acknowledgments

This work was supported by the National Science Foundation Graduate Research Fellowship, the Google Award for Inclusive Research, and the Social Science Research Council Just Tech Fellowship. The first author gratefully acknowledges the commitment of these organizations to advancing research at the intersection of technology equity and social justice.

# Ethical Considerations

This review synthesizes interdisciplinary scholarship on ASR, fairness, and linguistic inclusion, with a focus on AAE. We acknowledge that this work is not neutral: it involves interpretive choices that shape which perspectives are elevated in a field where Black voices and linguistic diversity are often marginalized.

Our ethical commitments are rooted in our positionality. (Harding 2004), as we advance care for underrepresented communities, epistemic humility across disciplines, and transparency in how we synthesize and critique scholarship. This work is led by Black researchers whose lived experiences and interdisciplinary expertise in HCI and sociotechnical AI inform a commitment to addressing systemic anti-Blackness and linguistic erasure.

# Acknowledgments

This work was supported by the National Science Foundation Graduate Research Fellowship, the Google Award for Inclusive Research, and the Social Science Research Council Just Tech Fellowship. The first author gratefully acknowledges the commitment of these organizations to advancing research at the intersection of technology equity and social justice.

# Appendices

## A. Governance-Centered ASR Life-cycle Framework

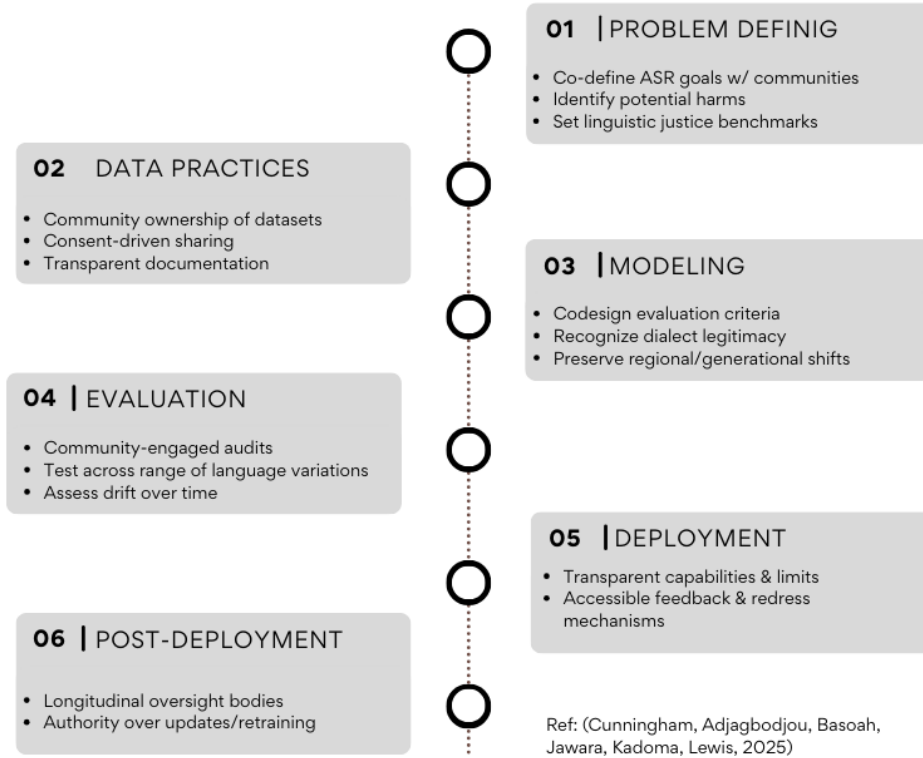

Across the literature, three patterns recur: (1) bias assessments tend to be narrowly technical, often relying on aggregate word error rates without deeper engagement with socio-technical harms; (2) data practices aimed at inclusion, while growing, are largely extractive and lack mechanisms for community consent, ownership, or control; and (3) participatory methods are underutilized, frequently limited to short-term consultation rather than sustained co-governance. These gaps indicate that fairness interventions remain fragmented and reactive, leaving structural power dynamics unaddressed. The governance-centered ASR life-cycle responds to these shortcomings by embedding participatory checkpoints, community agency, and accountability mechanisms at every stage of development, transforming isolated technical fixes into continuous, shared governance of speech AI systems.

## B. Expanded Details on Information Sources and Search Strategy

While we did not systematically track the precise frequency or inclusion rate of each keyword, the matrix served as an iterative tool for refining search strings and ensuring breadth of coverage. From this larger pool, we distilled a final list of focused keywords that were used in systematic database queries across ACM Digital Library, IEEE Xplore, ACL Anthology, Linguistics and Language Behavior Abstracts (LBA). In addition, we expanded our search to include open-source library sources such as arXiv, Frontiers, and PNAS. Representative terms from the final queries included "ASR fairness," "racial bias in speech recognition," "AAE speech errors," "linguistic bias in NLP," "participatory design ASR," and "inclusive speech technology."

Table 2: Summary of Keyword Clusters, Seed Terms, and Representative Expanded Terms Used in the Literature Search. The full keyword expansion generated 1,872 terms; representative terms are shown for brevity. Explanation of this can be found in the Methods section.

| Cluster | Seed Keywords | Representative Expanded Terms |
| --- | --- | --- |
| Responsible Methods & Design Approaches | algorithmic fairness, participatory design | responsible AI methods, human-centered AI, equitable machine learning, inclusive design frameworks |
| Social & Algorithmic Justice Challenges | bias in speech technology, racial bias | socio-technical harms, justice in NLP, discrimination in AI, algorithmic harm mitigation |
| Speech & Language Technologies (ASR Focus) | ASR fairness, speech recognition accuracy | voice bias detection, inclusive ASR evaluation, dialect-aware ASR models, speech-to-text performance |
| Language Varieties & Dialects (U.S. Context) | African American English, dialect recognition | sociophonetic variation, vernacular speech patterns, dialect-sensitive NLP, phonological variation |
| Racial & Ethnic Communities | Black speech patterns, minoritized language | heritage language technologies, underrepresented communities in AI, linguistic diversity preservation |
| Language Group Associations | language advocacy, speech community rights | linguistic rights in AI, cultural language preservation, community-led speech datasets |

## C. Reviewed Literature by Citation, Discipline, and Source

*The full table of literature can be found on the following 3-pages. Readers should note that the table appear as figures due to technical rendering limitations.*

Table 2: Reviewed Studies by Citation, Discipline, and Source

| Citation | Authors | Title | Discipline | Study Type | Info Source |
|---|---|---|---|---|---|
| (Sutton et al. 2019) | Sutton, Selina Jeanne; Foulkes, Paul; Kirk, David; Lawson, Shaun | Voice as a Design Material: Sociophonetic Inspired Design Strategies in Human-Computer Interaction | HCI | Empirical | ACM Digital Library |
| (Iacobelli and Cassell 2007) | Iacobelli, Francisco; Cassell, Justine | Ethnic Identity and Engagement in Embodied Conversational Agents | HCI | Empirical | ACM Digital Library |
| (Nkemelu et al. 2023) | Nkemelu, Daniel; Shah, Harshil; Essa, Irfan; Best, Michael L. | Tackling Hate Speech in Low-resource Languages with Context Experts | HCI | Empirical | ACM Digital Library |
| (Papakyriakopoulos and Xiang 2023) | Papakyriakopoulos, Orestis; Xiang, Alice | Considerations for Ethical Speech Recognition Datasets | HCI | Empirical | ACM Digital Library |
| (Hutiri and Ding 2022) | Hutiri, Wiebke Toussaint; Ding, Aaron Yi | Bias in Automated Speaker Recognition | HCI | Empirical | ACM Digital Library |
| (Dixon et al. 2018) | Dixon, Lucas; Li, John; Sorensen, Jeffrey; Thain, Nithum; Vasserman, Lucy | Measuring and Mitigating Unintended Bias in Text Classification | HCI | Empirical | ACM Digital Library |
| (Fabio et al. 2021) | Fabio, Poletto; Valerio, Basile; Manuela, Sanguinetti; Bosco, Cristina; Viviana, Patti | Resources and benchmark corpora for hate speech detection: a systematic review | HCI | Empirical | ACM Digital Library |
| (Chan and Rosenfeld 2012) | Chan, Hao Yee; Rosenfeld, Roni | Discriminative pronunciation learning for speech recognition for resource scarce languages | HCI | Empirical | ACM Digital Library |
| (Kim et al. 2022) | Kim, Junhan; Muhic, Jana; Robert, Lionel Peter; Park, Sun Young | Designing Chatbots with Black Americans with Chronic Conditions: Overcoming Challenges against COVID-19 | HCI | Empirical | ACM Digital Library |
| (Garg et al. 2023) | Garg, Tanmay; Masud, Sarah; Suresh, Tharun; Chakraborty, Tanmoy | Handling Bias in Toxic Speech Detection: A Survey | HCI | Empirical | ACM Digital Library |
| (Wenzel and Kaufman 2023) | Wenzel, Kimi; Kaufman, Geoff | Challenges in Designing Racially Inclusive Language Technologies | HCI | Conceptual | ACM Digital Library |
| (Harrington et al. 2022) | Harrington, Christina N.; Garg, Radhika; Woodward, Amanda; Williams, Dimitri | Its Kind of Like Code-switching: Black Older Adults Experiences with a Voice Assistant for Health Information Seeking | HCI | Empirical | ACM Digital Library |
| (Wenzel et al. 2023) | Wenzel, Kimi, Nitya Devireddy, Cam Davison, and Geoff Kaufman | Can Voice Assistants Be Microaggressors? Cross-Race Psychological Responses to Failures of Automatic Speech Recognition | HCI | Empirical | ACM Digital Library |
| (Lister et al. 2020) | Lister, Kate; Coughlan, Tim; Iniesto, Francisco; Freear, Nick; Devine, Peter | Accessible conversational user interfaces: considerations for design | HCI | Empirical | ACM Digital Library |
| (Reitmaier et al. 2022) | Reitmaier, Thomas; Wallington, Electra; Kalarikalayil Raju, Dani; Klejch, Ondrej; Pearson, Jennifer; Jones, Matt; Bell, Peter; Robinson, Simon | Opportunities and Challenges of Automatic Speech Recognition Systems for Low-Resource Language Speakers | HCI | Empirical | ACM Digital Library |
| (Markl and McNulty 2022) | Markl, Nina | Language variation and algorithmic bias: understanding algorithmic bias in British English automatic speech recognition | HCI | Empirical | ACM Digital Library |
| (Cunningham 2023) | Cunningham, Jay L. | Collaboratively Mitigating Racial Disparities in Automated Speech Recognition and Language Technologies with African American English Speakers: Community-Collaborative and Equity-Centered Approaches Toward Designing Inclusive Natural Language Systems | HCI | Empirical | ACM Digital Library |

Table of Citations displaying the authors, title of the piece, discipline of the published work, the type of study incorporated, and where the source was retrieved from.

| Citation | Authors | Title | Discipline | Study Type | Info Source |
| --- | --- | --- | --- | --- | --- |
| (Xu et al. 2021) | Xu, Albert; Pathak, Eshaan; Wallace, Eric; Gururangan, Suchin; Sap, Maarten; Klein, Dan | Detoxifying Language Models Risks Marginalizing Minority Voices | AI/ML | Empirical | ACLWeb |
| (Shah, Schwartz, and Hovy 2020) | Shah, Deven Santosh; Schwartz, H. Andrew; Hovy, Dirk | Predictive Biases in Natural Language Processing Models: A Conceptual Framework and Overview | HCI | Conceptual | ACLWeb |
| (Joshi et al. 2020) | Joshi, Pratik; Santy, Sebastin; Budhiraja, Amar; Bali, Kalika; Choudhury, Monojit | The State and Fate of Linguistic Diversity and Inclusion in the NLP World | HCI | Empirical | ACLWeb |
| (Sheng et al. 2021) | Sheng, Emily; Chang, Kai-Wei; Natarajan, Premkumar; Peng, Nanyun | Societal Biases in Language Generation: Progress and Challenges | Linguistics | Empirical | ACLWeb |
| (Bender and Friedman 2018) | Bender, Emily M.; Friedman, Batya | Data Statements for Natural Language Processing: Toward Mitigating System Bias and Enabling Better Science | HCI | Empirical | ACLWeb |
| (Bhatt et al. 2022) | Bhatt, Shaily; Dev, Sunipa; Talukdar, Partha; Dave, Shachi; Prabhakaran, Vinodkumar | Cultural Re-contextualization of Fairness Research in Language Technologies in India | HCI | Empirical | ACLWeb |
| (Sap et al. 2019) | Sap, Maarten; Card, Dallas; Gabriel, Saadia; Choi, Yejin; Smith, Noah A. | The Risk of Racial Bias in Hate Speech Detection | HCI | Empirical | ACLWeb |
| (Markl and McNulty 2022) | Markl, Nina; McNulty, Stephen Joseph | Language technology practitioners as language managers: arbitrating data bias and predictive bias in ASR | HCI | Conceptual | ACLWeb |
| (Jørgensen, Hovy, and Søgaard 2015) | Jørgensen, Anna; Hovy, Dirk; Søgaard, Anders | Challenges of studying and processing dialects in social media | HCI | Empirical | ACLWeb |
| (Blodgett et al. 2020) | Blodgett, Su Lin; Barocas, Solon; Daumé III, Hal; Wallach, Hanna | Language (Technology) is Power: A Critical Survey of "Bias" in NLP | HCI | Empirical | ACLWeb |
| (Dorn 2019) | Dorn, Rachel | Dialect-Specific Models for Automatic Speech Recognition of African American Vernacular English | HCI | Empirical | ACLWeb |
| (Ziems et al. 2022) | Ziems, Caleb; Chen, Jiaao; Harris, Camille; Anderson, Jessica; Yang, Diyi | VALUE: Understanding Dialect Disparity in NLU | HCI | Empirical | ACLWeb |
| (Noble and Bernardy 2022) | Noble, Bill; Bernardy, Jean-philippe | Conditional Language Models for Community-Level Linguistic Variation | HCI | Empirical | ACLWeb |
| (Davidson, Bhattacharya, and Weber 2019) | Davidson, Thomas; Bhattacharya, Debasmita; Weber, Ingmar | Racial Bias in Hate Speech and Abusive Language Detection Datasets | HCI | Empirical | ACLWeb |

Table of Citations displaying the authors, title of the piece, discipline of the published work, the type of study incorporated, and where the source was retrieved from.

| Citation | Authors | Title | Discipline | Study Type | Info Source |
|---|---|---|---|---|---|
| (Lhoest et al. 2021) | Lhoest, Quentin; Villanova del Moral, Albert; Jernite, Yacine; Thakur, Abhishek; von Platen, Patrick; Patil, Suraj; Chaumond, Julien; Drame, Mariama; Plu, Julien; Tunstall, Lewis; Davison, Joe; Šaško, Mario; Chhablani, Gunjan; Malik, Bhavitvya; Brandeis, Simon; Le Scao, Teven; Sanh, Victor; Xu, Canwen; Patry, Nicolas; McMillan-Major, Angelina; Schmid, Philipp; Gugger, Sylvain; Delangue, Clément; Matussière, Théo; Debut, Lysandre; Bekman, Stas; Cistac, Pierric; Goehringer, Thibault; Mustar, Victor; Lagunas, François; Rush, Alexander; Wolf, Thomas | Datasets: A Community Library for Natural Language Processing | HCI | Empirical | ACLWeb |
| (Blodgett, Green, and O'Connor 2016) | Blodgett, Su Lin; Green, Lisa; O'Connor, Brendan | Demographic Dialectal Variation in Social Media: A Case Study of African-American English | Linguistics | Empirical | ACLWeb |
| (Manzini et al. 2019) | Manzini, Thomas; Yao Chong, Lim; Black, Alan W; Tsvetkov, Yulia | Black is to Criminal as Caucasian is to Police: Detecting and Removing Multiclass Bias in Word Embeddings | HCI | Empirical | ACLWeb |
| (Santiago et al. 2022) | Santiago, Harrison; Martin, Joshua; Moeller, Sarah; Tang, Kevin | Disambiguation of morpho-syntactic features of African American English – the case of habitual be | HCI | Empirical | ACLWeb |
| (Groenwold et al. 2020) | Groenwold, Sophie; Ou, Lily; Parekh, Aesha; Honnavalli, Samhita; Levy, Sharon; Mirza, Diba; Wang, William Yang | Investigating African-American Vernacular English in Transformer-Based Text Generation | HCI | Empirical | ACLWeb |
| (Bird 2024) | Bird, Steven | Decolonising Speech and Language Technology | HCI | Conceptual | ACLWeb |
| (Koenecke et al. 2020) | Koenecke, Allison; Nam, Andrew; Lake, Emily; Nudell, Joe; Quartey, Minnie; Mengesha, Zion; Toups, Connor; Rickford, John R.; Jurafsky, Dan; Goel, Sharad | Racial disparities in automated speech recognition | HCI | Empirical | pnas.org (Atypon) |
| (Mengesha et al. 2021) | Mengesha, Zion; Heldreth, Courtney; Lahav, Michal; Sublewski, Juliana; Tuennerman, Elyse | "I don't Think These Devices are Very Culturally Sensitive."—Impact of Automated Speech Recognition Errors on African Americans | HCI | Empirical | Frontiers |
| (Prabhakaran, Qadri, and Hutchinson 2022)(Prabhakaran et al., n.d.) | Prabhakaran, Vinodkumar; Qadri, Rida; Hutchinson, Ben | Cultural Incongruencies in Artificial Intelligence | HCI | Empirical | arXiv.org |
| (Sarı, Hasegawa-Johnson, and Yoo 2021) | Sarı, Leda; Hasegawa-Johnson, Mark; Yoo, Chang D. | Counterfactually Fair Automatic Speech Recognition | HCI | Empirical | IEEE Xplore |
| (Thomason 2021) | Thomason, Sarah G. | 14 Managing Sociolinguistic Data with the Corpus of Regional African American Language (CORAAL) | HCI | Empirical | IEEE Xplore |
| (Johnson et al. 2022)) | Johnson, Alexander; Fan, Ruchao; Morris, Robin; Alwan, Abeer | LPC Augment: an LPC-based ASR Data Augmentation Algorithm for Low and Zero-Resource Children's Dialects | HCI | Empirical | IEEE Xplore |

| Citation | Authors | Title | Discipline | Study Type | Info Source |
|---|---|---|---|---|---|
| (Okpala et al. 2022) | Okpala, Ebuka; Cheng, Long; Mbwambo, Nicodemus; Luo, Feng | AAEBERT: Debiasing BERT-based Hate Speech Detection Models via Adversarial Learning | HCI | Empirical | IEEE Xplore |

Figure 3: Reviewed Literature by Citation, Discipline, and Source

Table of Citations displaying the authors, title of the piece, discipline of the published work, the type of study incorporated, and where the source was retrieved from.